\documentclass[numberedappendix,twocolumn,onecolappendix]{openjournal}

\usepackage{newtxtext,newtxmath}

\usepackage[T1]{fontenc}

\DeclareRobustCommand{\VAN}[3]{#2}
\let\VANthebibliography\thebibliography
\def\thebibliography{\DeclareRobustCommand{\VAN}[3]{##3}\VANthebibliography}

\usepackage{graphicx}	
\usepackage{amsmath}	
\usepackage{lipsum}
\usepackage[dvipsnames]{xcolor}
\usepackage{subfigure}
\usepackage{graphicx}
\usepackage{hyperref}	
\hypersetup{colorlinks=true,linkcolor=blue,citecolor=blue,filecolor=blue,urlcolor=blue}
\usepackage[export]{adjustbox} 

\newcommand{\Msun}{M_\odot}

\begin{document}
\title[Machine Learning Halo Properties I]{Machine Learning the Dark Matter Halo Mass of Milky Way-Like Systems}

\author{Elaheh Hayati$^{1,*,\dag}$}

\author{Peter Behroozi$^{1,2}$}

\author{Ekta Patel$^{3,4,\ddag}$}

\affiliation{$^{1}$Department of Astronomy and Steward Observatory, University of Arizona, 933 N Cherry Ave, Tucson, AZ, 85721, USA}

\affiliation{$^{2}$Division of Science, National Astronomical Observatory of Japan, 2-21-1 Osawa, Mitaka, Tokyo 181-8588, Japan}

\affiliation{$^{3}$Department of Astronomy, University of California, Berkeley, 501 Campbell Hall, Berkeley, CA, 94720, USA}

\affiliation{$^{4}$Department of Physics and Astronomy, University of Utah, 115 South 1400 East, Salt Lake City, Utah 84112, USA}

\thanks{* E-mail: ehayati@arizona.edu}
\thanks{$\dag$ LSSTC DSFP Fellow}
\thanks{$\ddag$ Hubble Fellow}

\shortauthors{Hayati et al.}



\begin{abstract}
Despite the Milky Way's proximity to us, our knowledge of its dark matter halo is fairly limited, and there is still considerable uncertainty in its halo mass. Many past techniques have been limited by assumptions such as the Galaxy being in dynamical equilibrium as well as nearby galaxies being true satellites of the Galaxy, and/or the need to find large samples of Milky Way analogs in simulations. 
Here, we propose a new technique based on neural networks that obtains high precision (<0.12 dex mass uncertainty with perfect measurements of 30 neighboring galaxies; <0.14 dex including fiducial observational errors) without assuming halo dynamical equilibrium or that neighboring galaxies are all satellites, and which can use information from a wide variety of simulated halos (even those dissimilar to the Milky Way) to improve its performance. This method uses only observable information including satellite orbits, distances to nearby larger halos, and the maximum circular velocity of the most massive satellite galaxy. In this paper, we demonstrate a proof-of-concept method on simulated dark matter halos; in future papers in this series, we will apply neural networks to estimate the masses of the Milky Way's and M31's dark matter halos, and we will train variations of these networks to estimate other halo properties including concentration, assembly history, and spin axis.
\end{abstract}




\section{Introduction}

In the current $\Lambda$CDM paradigm, dark matter is the dominant type of matter. For example, we expect that the Milky Way is surrounded by a dark matter halo that makes up most of its total mass. Because dark matter is not visible, it has been difficult to directly measure this mass around the Milky Way (MW), and hence there have been many studies that have attempted to estimate the Milky Way’s dark matter content via other means \citep[e.g.,][]{Oort26, Morrison00,Yanny00,Battaglia5, Frinchaboy8,Li08,Busha11, vanderMarel12,king15,Lowing15,Patel17,McMillan17,Patel18}. 

Recently, \citet{Wang20} reviewed the most common techniques that have been used to measure the Milky Way's halo mass, which we summarize here:
\begin{enumerate}
\item \textit{Estimating the Galactic escape velocity using high-velocity objects}: High-velocity stars 
do not remain in the Milky Way’s potential well for a long time, and therefore the velocity distribution of MW stars rapidly decreases above the escape velocity. Since the escape velocity is related to the halo mass profile, it is then possible to estimate halo mass from the measured stellar velocity distribution \citep[e.g.,][]{2007MNRAS.379..755S,2014A&A...562A..91P,2017MNRAS.468.2359W,2018A&A...616L...9M,2019MNRAS.485.3514D,2019MNRAS.487L..72G}.
\item \textit{Measuring the rotation curve
}: Circular velocities can be measured for gas in the interstellar medium (ISM) as well as maser sources and disk stars. In dynamical equilibrium, these are related to the enclosed mass via $M_\mathrm{enc}\propto V^{2}{R/G}$, with the constant of proportionality dependent on the assumed asphericity of the mass distribution
\citep[e.g.,][]{2002ApJ...573..597K,2011MNRAS.414.2446M,2012MNRAS.423.1109P,2013A&A...549A.137I,2017MNRAS.465...76M,2013JCAP...07..016N,2020MNRAS.494.4291C}. 
\item \textit{Modeling tracers (halo stars, globular clusters, and satellite galaxies) with the Spherical Jeans equation}: For regions beyond the Galactic disk, one can measure the radial velocity dispersion and velocity anisotropy of tracers and infer the enclosed mass using the Jeans equation. This method requires an assumption for the density profile, which has been determined to have a power-law form locally; this form is typically assumed valid to very large distances. The radial velocity dispersion is often measured observationally by assuming that it is the same as the line-of-sight velocity dispersion. The velocity anisotropy is determined by proper motion measurements of the tracers, which is a key uncertainty in this method since it is difficult to obtain high-quality proper motion data for tracers at large distances 
\citep[e.g.,][]{2005MNRAS.364..433B,2006MNRAS.369.1688D,2008ApJ...684.1143X,2010MNRAS.406..264W,2010ApJ...720L.108G,2014ApJ...785...63B,2016MNRAS.463.2623H,2017ApJ...846...10A,2018ApJ...862...52S,2018RAA....18..113Z,2020MNRAS.494.5178F}.
\item \textit{Modeling tracers (halo stars, globular clusters, and satellite galaxies) with phase-space distribution functions}: Using the assumption of steady state structure as well as an assumption about the shape of the potential, one can calculate phase-space distribution functions, i.e., the observed distributions of orbital energy and angular momentum for tracers of the potential. Via forward modeling of the true observations, it is then possible to reverse this process to infer the underlying gravitational potential well and the halo mass
\citep[e.g.,][]{1989ApJ...345..759Z,1996ApJ...457..228K,1999MNRAS.310..645W,2003A&A...397..899S,2012MNRAS.424L..44D,2015ApJ...806...54E,2017ApJ...835..167E,2019ApJ...875..159E}.
\item \textit{Simulating and modeling the dynamics of stellar streams}: Stellar stream shapes around the Galaxy provide information about galactic evolution and the underlying gravitational potential. The path of the stream and the different orbital speeds of objects along the streams tell us about the tidal forces that the object experienced, which can then be related to the potential well shape and the halo mass 
\citep[e.g.,][]{1995ApJ...439..652L,2005ApJ...619..807L,2010ApJ...711...32N,2014MNRAS.445.3788G,2015AAS...22514219K,2018AAS...23141204H,2019MNRAS.486.2995M,2019MNRAS.487.2685E}.
\item \textit{Modeling the motion of the Milky Way, M31, and other distant satellites under the framework of the Local Group timing argument}: Despite the expansion of the Universe, Andromeda and the Milky Way are approaching each other because of their gravitational pull. Under the assumption that the two galaxies are in a Keplerian orbit, one may infer their total mass by measuring other orbital properties including their relative velocity, their distance, and the age of the Universe 
\citep[e.g.,][]{1959ApJ...130..705K,1989ApJ...345..759Z,2008MNRAS.384.1459L,2012ApJ...753....8V,2013ApJ...768..139S,2020ApJ...888..114Z,2020ApJ...890...27Z,2023ApJ...942...18C}.
\item \textit{Measurements made by linking the brightest Galactic satellites to their counterparts in simulations}: In this method, one uses a Bayesian framework to measure the mass of the Milky Way by selecting simulated halos (i.e., from a dark matter simulation) that have satellites that are most similar to the satellites of the Milky Way. To select the best matches, it is important to have the proper motion of the satellites, as it has been shown that specific angular momentum is often a better constraint than knowing only the position, radial velocity, or orbital energy of the satellites \citep[e.g.,][]{2011ApJ...743..117B,2014MNRAS.445.2049C,Patel17,2017ApJ...850..116L,Patel18}.
\end{enumerate}
Each of the above techniques requires assumptions, which contribute to systematic uncertainties in constraining the Milky Way's dark matter halo mass. Most of the techniques above assume dynamical equilibrium for the Milky Way's halo. Dynamical equilibrium is known to be violated at small radii due to the passage of the Large Magellanic Cloud \citep[e.g.,][]{Laporte18,Garavito19} near the center of the Milky Way, and at large radii by continued accretion onto the halo \citep[e.g.,][]{McBride09,Behroozi13}. Nonetheless, dynamical equilibrium techniques share a strength that observations from arbitrary numbers of tracers can be combined.  

The techniques that do not assume dynamical equilibrium rely on $\Lambda$CDM simulations. While these methods can be designed to avoid systematic biases from out-of-equilibrium systems, they are limited in the amount of data they can combine: the more observational data one has, the more difficult it is to find simulated halos that match all the observational constraints simultaneously \citep[see discussion in, e.g.,][]{Patel18}.

Here, we use a new approach for measuring the Milky Way’s dark matter halo mass. We train a neural network on simulated galaxies to learn the transformation for linking observable galaxy properties (starting with the specific angular momenta of satellites of the Milky Way) to halo masses. This method has the following benefits:
\begin{enumerate}
  \item No dynamical equilibrium assumptions are made.
  \item No assumptions about most nearby galaxies being satellites are made.
  \item The approach can learn about relationships between observables (e.g., satellite orbits) and mass even from halos that do not match the MW or M31, leading to greater constraining power.
  \item Arbitrary constraints from the local or larger-scale environment (e.g., distance and/or velocity offsets to the nearest larger halo) can be self-consistently included.
\end{enumerate}
This paper is the first in a series that will explore the ability of neural networks to constrain the properties of the Local Group's dark matter distribution. The goal in this paper is to explore how well the technique performs before adapting it to the Milky Way or M{31} and their satellite systems. In the appendix, we consider a generic error model that is independent of the sky position where satellites are detected. While beyond the scope of the current paper, neural networks in the future will also provide the advantages of:
\begin{enumerate}
  \item Being able to use arbitrary non-dark matter tracers (e.g., gas rotation curves in hydrodynamical simulations) as input features to neural networks to achieve the most accurate mass constraints.
  \item Being able to use domain adaptation techniques \citep[e.g.,][]{Ciprijanovic22} to identify mass-observable relationships that are independent of baryonic physics differences across hydrodynamical simulations. 
  \item Being able to estimate other halo properties as well, just by changing the training target to other halo properties. Such properties could include the halo spin axis, halo concentration, and halo mass assembly history, with minimal additional effort.
\end{enumerate}

In this paper, we use dark matter halo simulations to train neural networks to estimate masses across a broad halo mass range ($10^8 - 10^{14}\, \Msun$). Inputs to the neural networks are based on observables including neighboring galaxy orbits, maximum circular velocity of the largest satellite, and distances to nearby more massive halos. In this paper, we take the limit of perfect information, assuming that no observational errors exist, as well as test the impact of a fiducial observational error model. In the second paper in this series, we will convolve simulated halo and galaxy properties with realistic observational errors, re-train the network, and use observed satellite orbits from \textit{Gaia} DR3 to estimate the mass of the Milky Way's and Andromeda's dark matter halos. In the third paper in this series, we will extend the analysis to predict Milky Way halo properties beyond mass, including concentration, spin axis, and assembly history.

This paper is organized as follows. In Section \ref{s:methods}, we describe the training process and dark matter simulations; in Section \ref{s:results}, we illustrate the performance of the resulting neural networks; and we discuss these results and provide conclusions in Section \ref{s:discussion}. Appendix \ref{a:errors} provides results including fiducial observational errors. We assume a flat, $\Lambda$CDM universe with $\Omega_{m} = 0.307$, $\Omega_{\Lambda} = 0.693$, $n_{s}=0.96$, $h=0.68$ and $\sigma_{8} =0.823$.
We adopt the virial halo mass definition $M_\mathrm{vir}$ from \citep{Bryan98}, i.e., the total mass (dark + baryonic) within a radius $R_\mathrm{vir}$ of a density peak.

\section{Methods}

\label{s:methods}

\subsection{Dark Matter Simulation}

For this work, we use the public Very Small MultiDark Planck (VSMDPL) simulation with $3840^{3}$ dark matter particles, each of mass $6.2 \times 10^{6}\Msun/h$. The simulation is based on a flat, $\Lambda$CDM universe with $\Omega_{m} = 0.307$, $\Omega_{\Lambda} = 0.693$, $n_{s}=0.96$, $h=0.68$ and $\sigma_{8} =0.823$. It evolves matter from $z=150$ to $z=0$ within a periodic cube of side length 160 comoving Mpc/$h$. There are 151 snapshots with identified halos between $z=0$ and $z=25$. Halos are identified using \textsc{Rockstar} \citep{Behroozi013}, and merger trees are identified using the \textsc{Consistent Trees} algorithm \citep{Behroozi2013}. Each halo is identified in the merger trees as a central halo or as a satellite halo (i.e., a halo contained within the virial radius of a larger halo). 
We adopt the virial halo mass definition $M_\mathrm{vir}$, i.e., the total mass (dark + baryonic) within a radius $R_\mathrm{vir}$ of a density peak, such that the average density enclosed is $\rho_\mathrm{vir}$ from \citet{Bryan98}.

\subsection{Intuition for Using Specific Angular Momenta}

One of the principal inputs to our neural networks is specific angular momenta of neighboring galaxies. Under our halo definition, both the halo radius and the halo circular velocity ($\sqrt{GM/R}$) scale as halo mass to the one-third power.

\begin{figure}
  \centering
  \includegraphics[width=1\columnwidth]{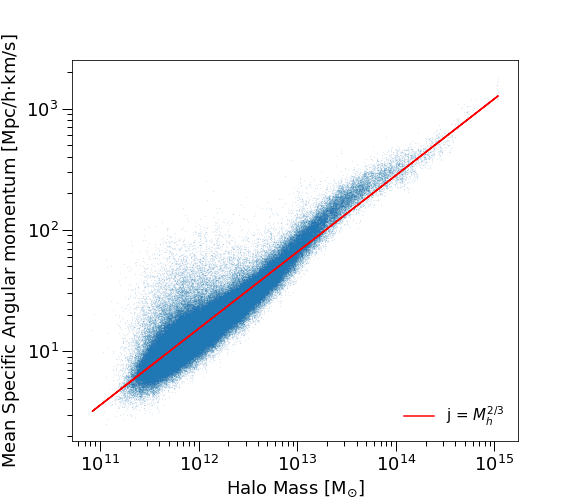}
  \caption{The average specific angular momenta of the 30 largest satellites (selected by highest peak $v_\mathrm{max}$) versus central halo mass, for dark matter halos in the VSMDPL simulation. The expected dependence on halo mass ($j \propto M_h^{2/3}$) is shown by the red line, which is generally tightly followed by the simulated halos.}
  \label{fig:Ang_mass}
\end{figure}

%
As a result, the characteristic distances and velocities of the satellite halos with respect to the host halo (which by dimensional analysis are proportional to the halo radius and circular velocity) both scale as host halo mass to the one-third power. The characteristic specific angular momenta of satellites then depends on halo mass to the two-thirds power:
\begin{equation}
  j=(R\times V) \propto R_\mathrm{vir} \times v_\mathrm{circ,vir} 
  \propto M_\mathrm{vir}^{2/3}.
\end{equation}
This characteristic scaling is evident across a broad mass range for all central halos in our simulation in Fig.\ \ref{fig:Ang_mass}, which demonstrates the average specific angular momenta of the 30 largest neighbors versus central halo mass.

As discussed in \citet{Patel18}, the specific angular momentum of the satellite galaxies provides strong constraints on host halo mass. As shown in Fig.\ \ref{fig:Ang_mass}, the $M_\mathrm{vir}^{2/3}$ scaling is evident for a very wide range of halo masses. Only halos above $10^{13.5}\Msun$ start to show a bend in the scaling relation, due to more radial orbits for massive halos. Additionally, halos below $M_\mathrm{vir} = 10^{12.5}\Msun$ show scatter towards high specific angular momenta, which occurs for lower-mass halos that are near much more massive halos.

In this paper, we do not assume advanced knowledge of which nearby galaxies are satellites and which are not. Nonetheless, satellite angular momenta are approximately conserved throughout their orbits \citep{Patel18}. Hence, even when bound and unbound galaxies are mixed in a given vicinity of a halo, the bound galaxies' orbits will appear as an overdensity in the specific angular momentum distribution of the neighboring galaxies, and so specific angular momenta still provide useful information about host halo mass.

\subsection{Halo Selection and Input Features}

To train our deep neural networks, we first select halos with peak masses (i.e., their largest historical halo mass) larger than $10{^8}\Msun$ from the VSMDPL simulation, as the simulation does not resolve lower-mass halos well. These are also the only halos expected to host galaxies for which proper motions can be measured, due to the atomic cooling limit suppressing star formation in lower-mass halos \citep[e.g.,][]{OShea15}. In contrast to past studies, we place no additional prior or selection on host halo masses, as this information comes from observables alone in our method.

Past studies to infer mass have typically assumed that all nearby galaxies are satellites of the Milky Way, which places a strong prior on host halo mass. Because we do not know this to be the case in reality, we drop this assumption in this study, instead using the orbital properties (including specific angular momentum $j$, radial distance $R$, and relative velocity $V$) of the largest neighboring halos out to a fixed distance as our main input features. For this paper, we select neighboring halos out to 200 kpc from central halos, corresponding approximately to the distance out to which proper motions can be measured for Milky Way satellite candidates with \textit{Gaia}.

In particular, we do not make any cuts on whether the neighbors are bound or not, as this information is not known \textit{a priori} from the observations. Past studies, including \citet{Patel18}, used the specific angular momenta of $\sim$10 satellites to infer the mass of the Milky Way's halo, whereas the \textit{Gaia} mission has now provided 6D phase space information (and therefore angular momenta) for $\sim$50 satellites \citep{Li21,Fritz18,McConnachie20} within 200 kpc. Hence, we train a 10-neighbor neural network (3,093,208 halos) to compare our approach with past approaches, and we also train a 30-neighbor neural network (222,612 halos)to show the improvement possible with our new approach.

In tests, we found that dropping the assumption of satellite membership made it very difficult for networks that used angular momenta alone to reliably estimate host halo mass. As discussed in later sections, the neighbors of low-mass halos ($<10^{11}\Msun$) do not have specific angular momentum distributions that correlate with halo mass; because low-mass halos are much more numerous than high-mass halos, training results in networks that try to limit the worst-case performance for low-mass halos, rather than improve the best-case performance for higher-mass halos. However, adding some observable information that correlates broadly with host halo mass can help networks discriminate between the cases where the specific angular momentum of neighboring halos correlates with halo mass and where it does not.

In this work, we use the maximum circular velocity, $v_\mathrm{max}$, of the most massive satellite (the Large Magellanic Cloud in the case of the Milky Way, or M33 in the case of Andromeda) to help the networks distinguish between whether they are in the low-mass (neighbor angular momenta uncorrelated with host halo mass) or high-mass (neighbor angular momenta correlated with host halo mass) regimes. Using $v_\mathrm{max}$ of the largest satellite in this way follows from past studies that have also done so \citep[see, e.g.,][]{Busha11,Patel17, Patel18, Patel2023}.

From Fig.\ \ref{fig:Ang_mass}, we know that nearby massive halos can influence the angular momentum distributions of satellites. Hence, we also include input features corresponding to the distance to the nearest larger halo ($D_\mathrm{larger}$) and the distance to the nearest larger halo with $M_\mathrm{vir}\ge 10^{14}\Msun$ ($D_\mathrm{14}$). At high mass, these quantities converge by definition.

\subsection{Network Training}

Neural networks consist of interconnected nodes organized into layers, and they are capable of learning intricate patterns and relationships from data. We have used a deep neural network (NN) for our regression task of estimating halo mass from galaxy observables. Deep NN's are commonly used for image- and language-related tasks, but they can also be applied to arbitrary structured data as in this paper.

The hyper-parameters and structure that we used in our neural networks are as follows:
\begin{enumerate}
\item Input Size: Our input layer has 3 features for the orbital properties ($j$ [specific angular momentum], $R$ [distance from halo center], $V$ [velocity offset from halo center]) of each neighboring halo, as well as an additional 3 features for the target halo's environment ($v_\mathrm{max}$ of most massive satellite, distance to nearest larger halo, and distance to nearest $10^{14}\Msun$ halo). For the 10-neighbor network, this totals 33 input features, and for the 30-neighbor network, this totals 93 input features.

\item Layer Architecture: We use 5 fully-connected hidden layers. Each hidden layer (i.e., a layer in between the input and output layers) contains neurons that apply a nonlinear transformation to the input features, which are taken from the outputs of the previous layer. Fully connected layers are those in which every neuron in a given layer receives an input from every neuron in the previous layer. Initially, we have 10 neurons in the first hidden layer. Progressing through the network, we decrease the number of neurons in each subsequent layer (8, 6, 4, 2). This is known as a decreasing architecture, and it helps in reducing the complexity of the information passed through each layer as we go deeper into the network. 

\item Activation Function: We have used Rectified Linear Unit (ReLU) activation functions in our hidden layers. ReLU is a common choice because it introduces non-linearity into the model while being computationally efficient. Nonlinearity is essential in neural networks--otherwise the action of the neural network could be represented by a linear transform (i.e., a matrix multiplication), which would prevent it from learning complex, nonlinear relationships between the input and output data.

\item Output Layer: We have a single neuron in the output layer, since our network is performing regression to predict a single output (i.e., the mass of a central halo).

\item Loss Function: We have chosen Mean Squared Error (MSE) as our loss function, i.e., the metric by which we judge the neural network's performance. MSE is commonly used for regression tasks and calculates the average of the squared differences between predicted and actual values. It penalizes larger errors more heavily.

\item Optimizer: We have chosen the Adam optimizer. Adam is an adaptive learning rate optimization algorithm that combines the benefits of two other popular optimizers, RMSprop and Momentum, in that it adaptively chooses how far to proceed along the gradient of the loss function for each update to the neural network parameters. It is well-suited for a wide range of problems and often converges faster than traditional stochastic gradient descent (SGD).

\item Learning Rate: Our learning rate is set to 0.001. This parameter controls the initial step size during optimization. The value of 0.001 is a common starting point, but its value can be tuned depending on the specific problem and data set.

\item Batch Size: Our batch size is 64. This determines the number of input data points used in each update of the neural network's weights during training. Smaller batch sizes can lead to noisier updates but are more computationally efficient, while larger batch sizes provide smoother updates but require more time to compute each update.
\end{enumerate}

For training, we select all central halos with at least $N$ neighbors within 200 kpc (with $N=10$ or $30$, as appropriate). As above, we place no prior on central halo mass, so these halos range from $\sim 10^8 - 10^{15}\Msun$. We use three orbital parameters ($j$, $R$, and $V$) for each of the $N$ neighbors with the highest peak $v_\mathrm{max}$ as inputs to the neural network, as a proxy for the brightest galaxies \citep{Reddick13}. We also use the $v_\mathrm{max}$ of the most massive satellite (corresponding to the $v_\mathrm{max}$ of the Large Magellanic Cloud for the Milky Way), the distance to the nearest larger halo (corresponding to the distance to M31 for the Milky Way), and the distance to the nearest $10^{14}\Msun$ or larger halo (corresponding to the Virgo Cluster for the Milky Way) as input parameters. As above, the 10-neighbor network has 33 input features, and the 30-neighbor has 93 input features.

We used simulation snapshots from $z=0$ to $z=0.25$ from the VSMDPL simulation to increase the diversity of neighboring halo orbital configurations available for training. We found that including training data from earlier snapshots did not cause a measurable bias in median predicted masses for $z=0$ halos, suggesting that the distribution of orbital configurations has not changed significantly over this redshift interval. Halos are split into a training sample (63\%) and a test sample (37\%) according to whether the halos have an X-coordinate less than or greater than $96$ Mpc/$h$ (compared to an overall box length of $160$ Mpc/$h$). This division is made to capture the uncertainties arising both from Poisson statistics and larger-scale cosmic variance.

\begin{figure}
  \centering
  \includegraphics[width=1\columnwidth]{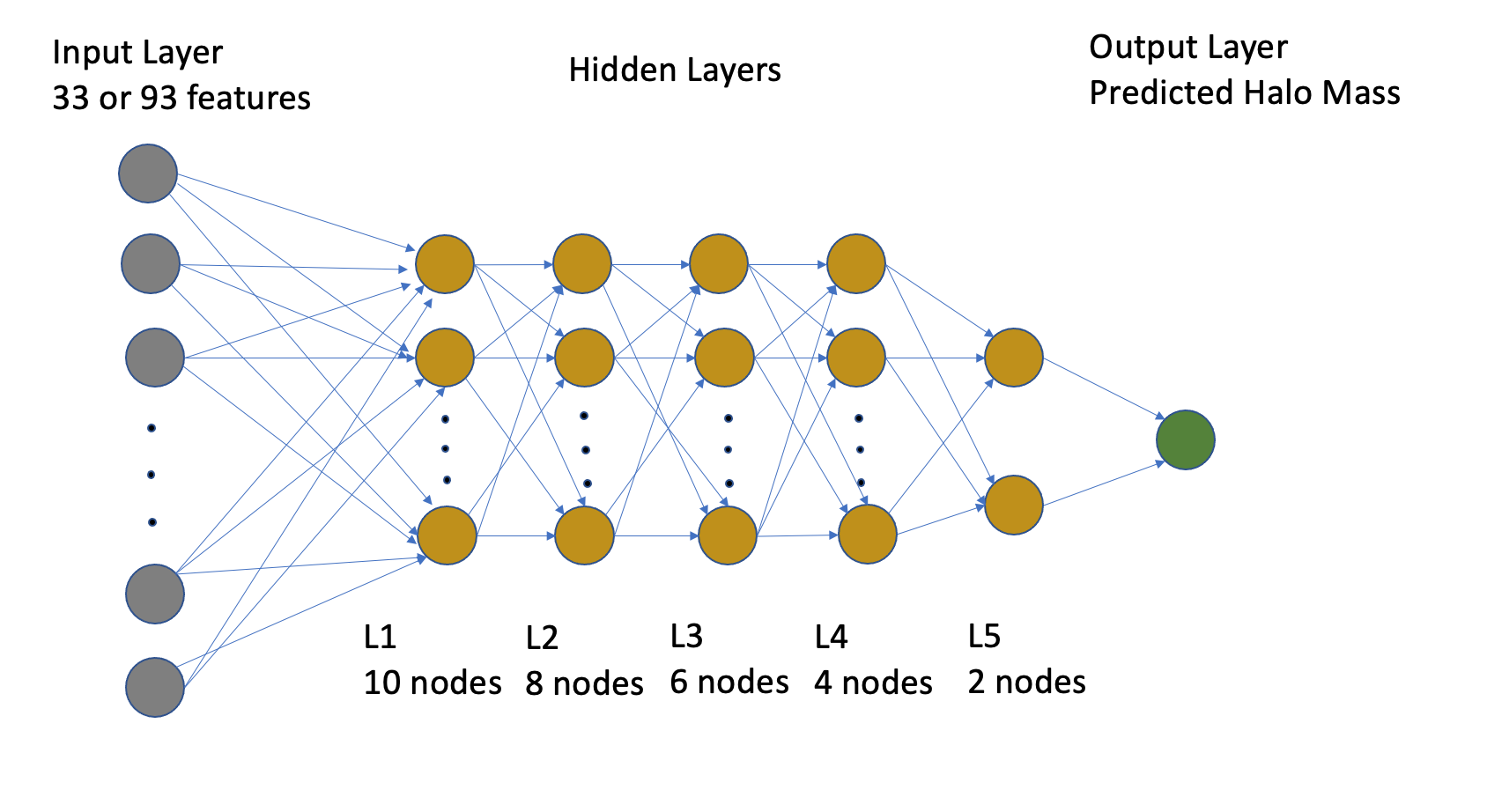}
  \caption{The neural network geometry we use to predict halo masses. Input features include neighboring halos' specific angular momenta ($j$), radial distances ($R$), and relative velocities ($V$), as well as the maximum circular velocity of the most massive satellite ($v_\mathrm{max,sat}$), the distance to the nearest larger halo ($D_\mathrm{larger}$), and the distance to the nearest halo with $M_\mathrm{vir}>10^{14}\Msun$ ($D_\mathrm{14}$). For all networks (regardless of the number of inputs), there are 5 hidden layers gradually decreasing from 10 nodes to 2 nodes, with one output layer corresponding to the predicted halo mass.}
  \label{fig:Neural}
\end{figure}

To pre-process, we ordered neighboring halos by increasing specific angular momenta, took the logarithms of all input features, subtracted the mean values across all neighbors, and scaled to unit variance. We then trained two 5-layer fully connected neural networks on the 10- and 30-neighbor input feature vectors to predict the masses of the corresponding central halos. The details of the network structure are shown in Fig.~\ref{fig:Neural}, and the details of the hyper-parameters are shown in Table \ref{HP}.

We varied several different hyper-parameters for the training process: the number of layers, the learning rate, the number of nodes per layer, the loss function, and the batch size.  We used a hand search to tune the learning rate, batch size, and loss function. For the rest of the hyper-parameters, we started with a simple network and increased the size until the mean-squared error did not improve further. 

We did not find any substantial improvements over the fiducial choice of parameters in Table \ref{HP}, and in some cases found worse performance. For example, when using optimizers such as RMSprop or Adagrad, we observed that the network exhibited a loss of prediction accuracy, particularly at the high mass end. This suggests that these optimizer choices may have gotten stuck in local minima, as performance for the vast majority of the halo sample (i.e., low mass halos) was prioritized over performance for high-mass halos.

\begin{table*}
 \centering
 \begin{tabular}{c|c}
 Hyperparameter & Choices\\
 \hline
 \hline
 Learning Rate & 0.00005, 0.0001, \textbf{0.001}, 0.1\\
 \hline
 Batch Size & 1,10,\textbf{64},1000\\
 \hline
 Optimizer & \textbf{Adam},Adammax,RMSprop,Adagrad\\
 \hline
 Loss Function & \textbf{mean squared error}, log loss, Exponential Linear Unit\\
 \hline
 Activation Function & \textbf{Rectified Linear Unit}, Sigmoid,Softmax,Parametric Rectified Linear Unit
 
 \end{tabular}
 \caption{The fiducial hyper-parameters used to train the network (in \textbf{bold}), as well as variations explored.}
 \label{HP}
\end{table*}

\section{Results}
\label{s:results}

\subsection{Performance of the neural network approach}

\begin{figure*}
  \begin{minipage}{0.5\textwidth}
    \centering
    \includegraphics[width=\linewidth]{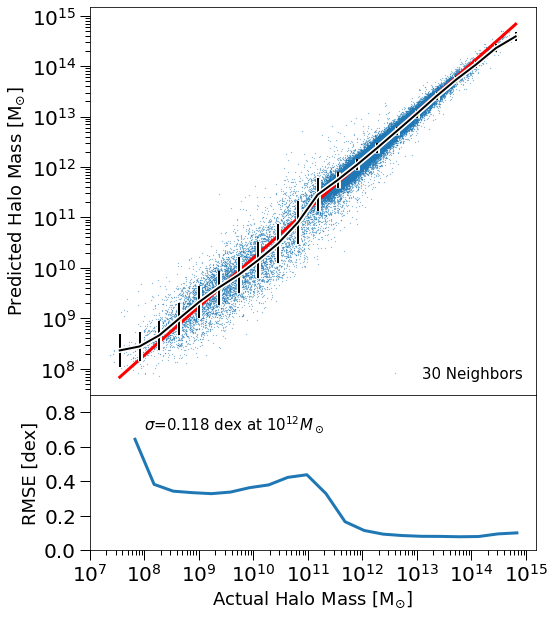}
  \end{minipage}%
  \begin{minipage}{0.5\textwidth}
    \centering
    \includegraphics[width=\linewidth]{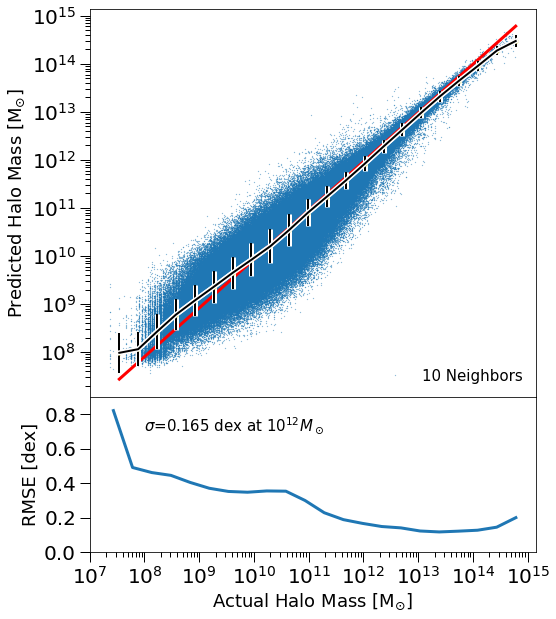}
  \end{minipage}
  \caption{Predicted halo mass versus actual halo mass for the neural networks in this paper applied to dark matter simulations. Input features to the networks correspond to observables, primarily including neighboring galaxies' specific angular momenta and other orbital properties. The left figure shows the result from halos with at least 30 neighboring galaxies, with reduced errors compared to the right figure, which used halos having at least 10 neighboring galaxies. In each figure, the bottom panels show the root mean square error (RMSE) as a function of actual halo mass. Typical errors are very good in both cases, about 0.17 dex for Milky Way-mass halos for the network using 10 neighbors and 0.12 dex for Milky Way-mass halos for the network using 30 neighbors. Error bars show the standard deviations of the predicted halo masses as a function of actual halo mass. The black line shows medians of predicted halo masses in bins of actual halo mass. The red line serves as a reference to indicate where the predicted mass would be equal to the actual mass.}
  \label{fig:overall}
\end{figure*}

We measure the performance of the neural network approach by applying the trained network to halos that it has never seen before (i.e., halos in our test set). The variance of the predicted halo masses at fixed actual halo mass then corresponds to the expected uncertainties of the network when applied to new data, such as for the Milky Way and M31. Hereafter, we quote network uncertainties at an actual halo mass of $10^{12}\Msun$ to represent the expected performance for the Milky Way and M31.

Fig.~\ref{fig:overall} summarizes the results of our work, demonstrating that the specific angular momenta of neighboring galaxies can be used to accurately infer the masses of central halos. The medians of the neural networks' predicted masses (in bins of actual halo mass) closely match actual halo masses, with typical median offsets of $\lesssim$ 0.03 dex at halo masses of $10^{12}\Msun$. However, the uncertainty in the predicted masses is significantly larger for low-mass halos (below a threshold of $\sim 10^{11.7}\Msun$) compared to high-mass halos. The size of the uncertainty is primarily influenced by whether the neighboring galaxies within 200 kpc are satellites or not. We investigate this aspect further in the next subsection.

The bottom plots in Fig.\ \ref{fig:overall}, show the RMS magnitudes of the errors across the full range of predicted masses. Specifically for MW-mass halos (again considering a threshold of $M_\mathrm{vir}\gtrsim 10^{11.7} \Msun$), the typical errors are $\sim$ 0.17 dex when using 10 neighboring halos, and they are $\sim$ 0.12  dex when using 30 neighboring halos, corresponding to a 30\% reduction in uncertainty. Since the ratio of these errors is less than expected from Poisson statistics ($0.17/0.12 \sim 1.4 < \sqrt{30/10} \sim 1.7$), this may be caused by correlated orbits known to occur in $\Lambda$CDM simulations, such as satellites coming in along the same filaments or even some satellites being satellites of other satellites 
\citep[e.g.,][]{2020ApJ...893..121P,2020MNRAS.498.5574E,2022A&A...657A..54B}.

\subsection{Understanding what information constrains halo masses}

\begin{figure*}[t!]
 \centering
  \includegraphics[width=\columnwidth]{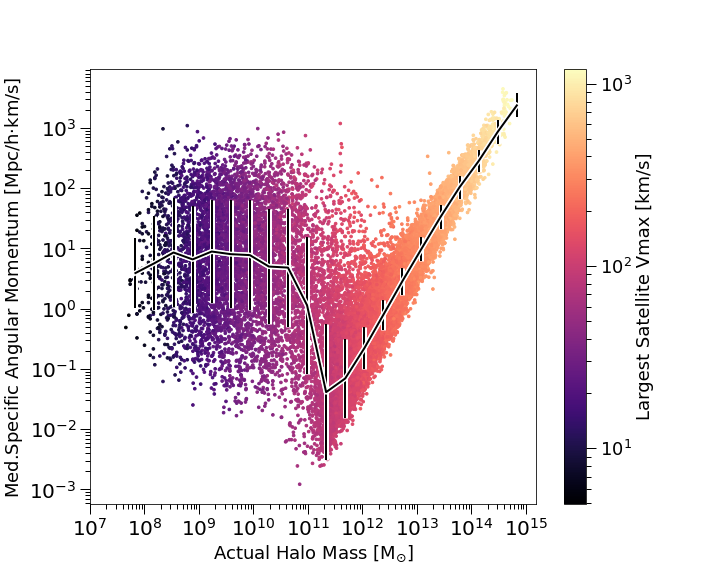} 
  \includegraphics[width=\columnwidth]{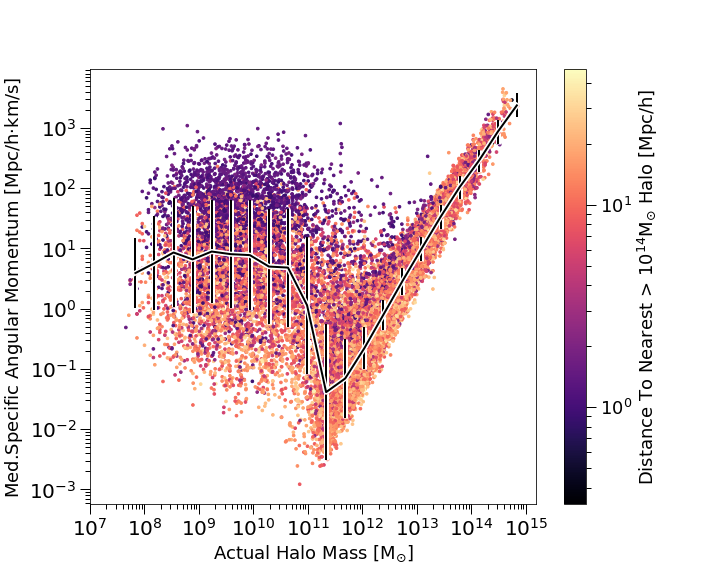}
  \caption{\textbf{Left}: the median specific angular momentum of neighboring halos as a function of halo mass, for halos that have at least 30 neighbors within 200 kpc. Halos are color-coded by the most massive satellite's maximum circular velocity, which correlates with host halo mass. Here, the neighbors of high-mass halos are much more likely to be satellites and thus have orbits with correlated specific angular momenta. In contrast, low-mass halos usually have non-satellite neighbors, which are less influenced by the low-mass halo's presence. So, the distributions of neighbors' specific angular momenta are much more correlated with halo mass for high-mass than low-mass halos. \textbf{Right}: the median specific angular momentum of halos' neighbors, now color-coded by the distance to the nearest massive halo ($M_h>10^{14}\Msun$). Gravitational forces from high-mass halos impact the orbits of all nearby halos, leading to higher relative velocities between low-mass halos and their neighbors. Moreover, massive halos have satellites that possess high velocities, and as these satellites' orbits can extend beyond the virial radii of the massive halos, they can pass nearby other lower-mass halos even as they have very large specific angular momentum offsets. Hence, the largest median specific angular momenta typically occur near massive halos.}
  
  \label{fig:median_J,J_larger}
 
\vspace{0.5cm} 

  \includegraphics[width=\columnwidth]{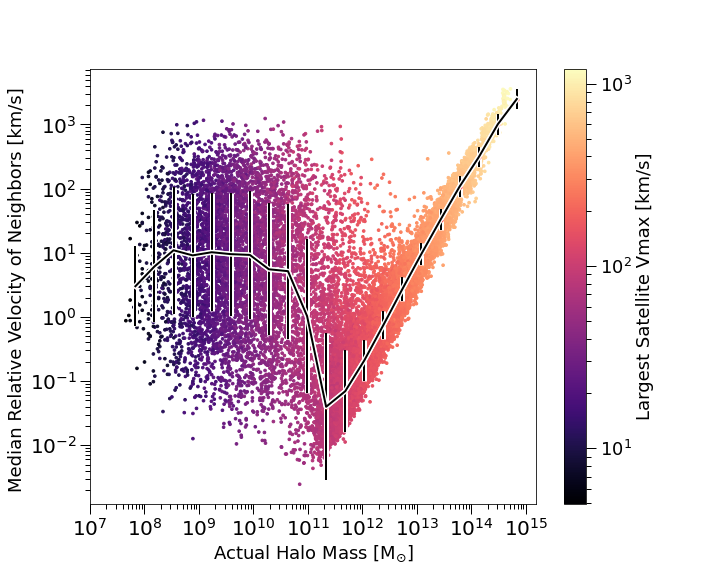}
  \includegraphics[width=\columnwidth]{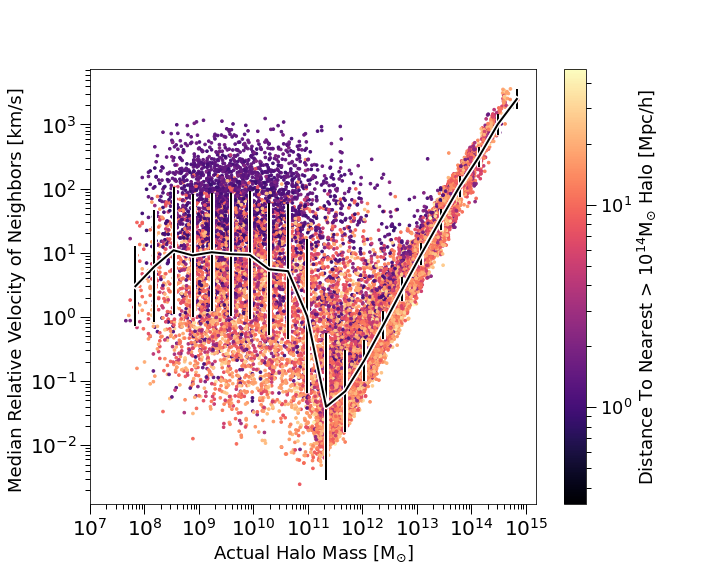}
  \caption{\textbf{Left}: the median relative velocities of neighboring halos as a function of halo mass, for those halos with 30 neighbors within 200 kpc from their centers. Halos are color-coded by the most massive satellite's maximum circular velocity, which correlates with host halo mass. Here, the neighbors of high-mass halos are much more likely to be satellites and thus have orbits with correlated relative velocities. In contrast, low-mass halos usually have non-satellite neighbors. As in Fig.\ \ref{fig:median_J,J_larger}, the distributions of neighbors' relative velocities are much more correlated with halo mass for high-mass than low-mass halos. \textbf{Right}: median relative velocities of halos' neighbors, now color-coded by the distance to the nearest massive halo ($M_h>10^{14}\Msun$). As in Fig.\ \ref{fig:median_J,J_larger}, the largest median neighbor relative velocities typically occur near massive halos.}
  
  \label{fig:median_V,V_larger}
\vspace{0.5cm} 
\end{figure*}

To analyze the relationships between satellite specific angular momenta ($j$), relative velocities ($V$), and radial distances ($R$) with respect to halo mass, we present Figures \ref{fig:median_J,J_larger} and \ref{fig:median_V,V_larger}. These figures illustrate the distributions of neighboring halos' orbital properties, where the left-hand panels are color-coded by the most massive satellite's maximum circular velocity ($v_{\mathrm{max,sat}}$), and the right-hand panels are color-coded by $D_\mathrm{14}$, the distance to the nearest massive halo ($M_\mathrm{vir}>10^{14} \mathrm{M}_{\odot}$).

The overall distributions of $j$, $R$, and $V$ exhibit distinct patterns, particularly with larger spreads observed for low-mass halos compared to high-mass halos. This can be attributed to the neighbors of high-mass halos being predominantly satellites of the high-mass halo, so the high-mass halo has a strong influence on its neighbors' orbits. However, neighbors of low-mass halos are typically not satellites and hence the presence of the low-mass halo does not strongly influence their orbits. Therefore, the distributions of neighbors' $j$ and $V$ are much more correlated with halo mass for high-mass than low-mass halos.

The color coding in the left-hand plots shows a smooth progression with actual halo mass, demonstrating a strong correlation between halo mass and the maximum radial velocity of the most massive satellite ($v_{\mathrm{max,sat}}$). Hence, the neural networks can effectively utilize satellite orbit information for large halo masses (when orbit information correlates with halo mass), while relying on the most massive satellite's maximum radial velocity as the best estimate when neighboring objects are not satellites.

The right-hand plots reveal that the presence of massive nearby halos biases neighbors' orbits, especially for low-mass halos. This outcome is expected since tidal forces from high-mass halos exert influence on the orbits of all neighboring halos, resulting in increased relative velocities between the low-mass halo and its neighbors. Additionally, massive halos have high satellite velocities, and because the orbits of satellite halos often extend beyond halos' virial radii (where they become known as ``backsplash'' or ``flyby'' halos; see, e.g., \citealt{Diemer21,ODonnell21}), some neighboring halos around low-mass halos will have orbits that are strongly influenced by their high-mass neighbors.

\begin{figure*}
 \centering
  \includegraphics[width=\columnwidth]{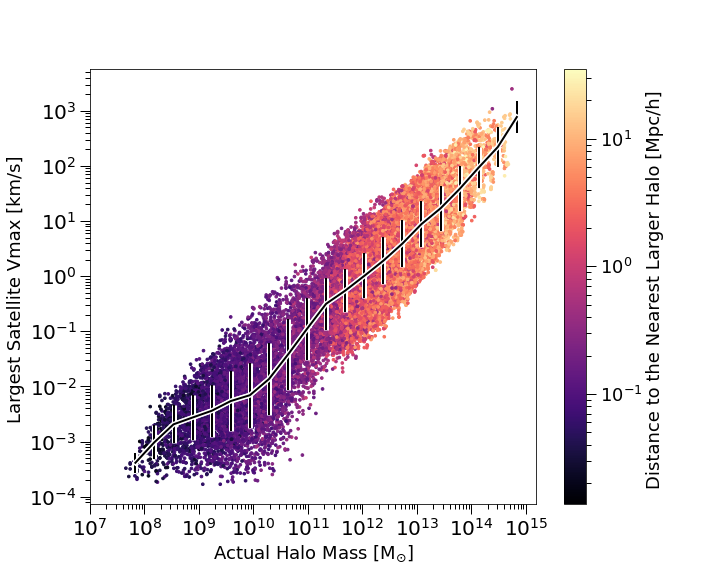} 
  \includegraphics[width=\columnwidth]{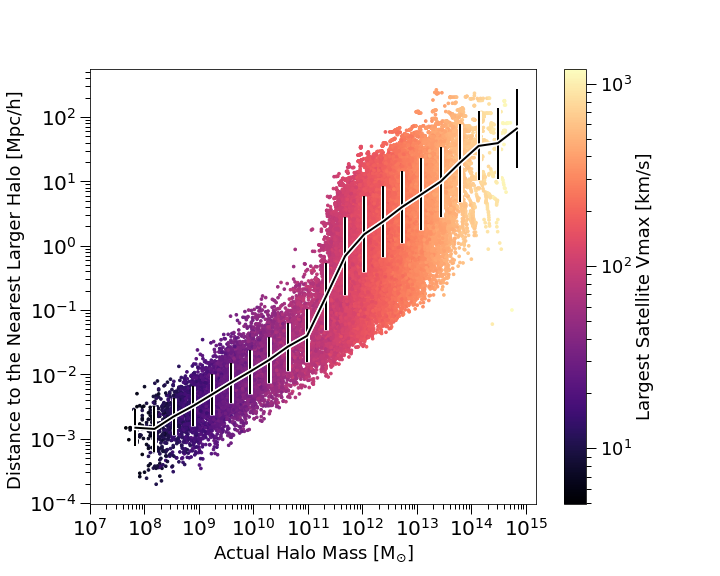}
  \caption{\textbf{Left}: There is a strong correlation between the maximum circular velocity of the most massive satellite and the host halo mass. 
 The color coding indicates the distance to the nearest larger halo, which is also correlated with host halo mass, but more weakly than the maximum circular velocity of the most massive satellite. \textbf{Right}: This plot shows the correlation between the distance to the nearest larger halo and the host halo mass. Larger halos are less prevalent, which leads to larger distances between them when compared to smaller halos. So, the distribution of larger halos contributes to a distinct pattern for $D_\mathrm{larger}$, different from that of $v_\mathrm{max,sat}$. There is a noticeable kink in the median relation between $D_\mathrm{larger}$ and halo mass around $M_\mathrm{vir}\sim 10^{11}\Msun$. This kink arises due to our selection criteria, where we focus on halos with a minimum of 30 neighbors within a 200 kpc radius.}
  
  \label{fig:median_Vmax,median_DNL}
 \end{figure*}
 
 \begin{figure*}
\centering
  \includegraphics[width=\columnwidth]{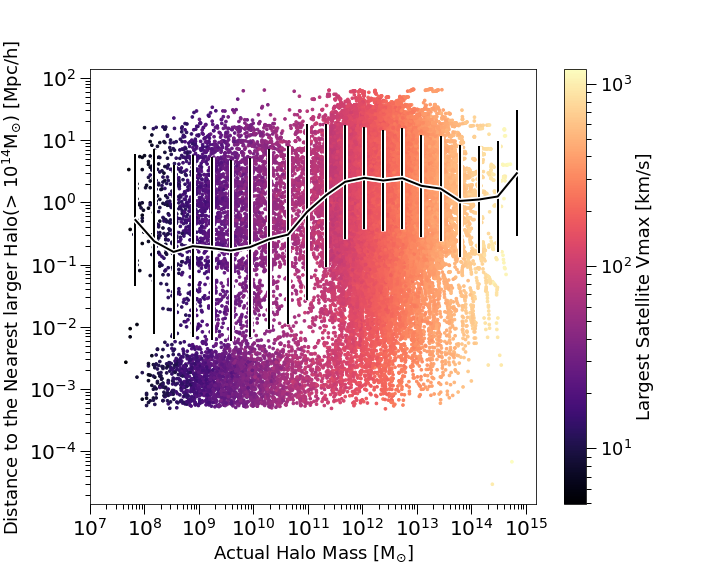}

  \caption{There is little correlation between the distance to the nearest massive halo and the host halo mass. Halos of all masses can be found near massive halos, which in turn can significantly impact orbital properties of their neighboring halos.}
  
  \label{fig:median_D14}

\end{figure*}

Figures~\ref{fig:median_Vmax,median_DNL} and \ref{fig:median_D14} show the relationship between three variables: distance to the nearest $M_\mathrm{vir}>10^{14}\Msun$ halo ($D_{14}$), distance to the nearest larger halo ($D_\mathrm{larger}$), and $v_\mathrm{max}$ of the most massive satellite ($v_\mathrm{max,sat}$), with respect to halo mass. The parameter $v_\mathrm{max,sat}$ has a very strong correlation with host halo mass, as larger halos typically host larger satellites. 

The parameter $D_\mathrm{larger}$ also exhibits a correlation with halo mass. However, the relationship is weaker and exhibits a different shape from that of $v_\mathrm{max,sat}$. Larger halos are relatively less common, which directly implies that the distances between large halos tend to be larger than the distances between small halos, despite the fact that larger halos are more biased relative to the underlying dark matter distribution. We also note that there is a kink in the median relation between $D_\mathrm{larger}$ and halo mass at $M_\mathrm{vir}\sim 10^{11}\Msun$, which occurs because we are selecting halos with at least 30 neighbors within a 200 kpc radius; for low-mass halos, this preferentially selects halos in dense environments, i.e., for which the distance to surrounding halos is significantly decreased.

Finally, unlike $v_\mathrm{max,sat}$ and $D_\mathrm{larger}$, $D_{14}$ does not exhibit much correlation with halo mass. This indicates that halos of varying masses are present across different environments, leading to a wide range of $D_{14}$ values irrespective of halo mass.

\begin{figure*}
    \begin{minipage}[t]{0.48\textwidth}
        \centering
        \includegraphics[width=\linewidth,valign=c]{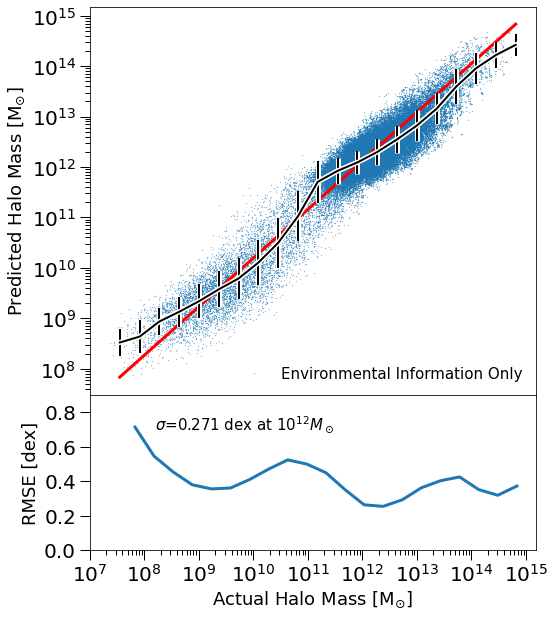}
        \caption{This figure shows a neural network trained on just three features, the distance to the nearest halo with $M_\mathrm{vir}>10^{14}\Msun$ ($D_{14}$), the distance to the nearest larger halo ($D_\mathrm{larger}$), and the maximum circular velocity of the most massive satellite ($v_\mathrm{max,sat}$). The uncertainties are now more similar across halo masses, with typical values of 0.271 dex at halo masses of $10^{12}\Msun$. This suggests that, while helpful, $v_\mathrm{max,sat}$ does not primarily determine halo mass, but instead most of the information leading to lower uncertainties in Fig.\ \ref{fig:overall} is coming from the orbits of neighboring halos.}
        \label{fig:3feat}
    \end{minipage}\qquad
    \begin{minipage}[t]{0.48\textwidth}
        \centering
        \includegraphics[width=\linewidth,valign=c]{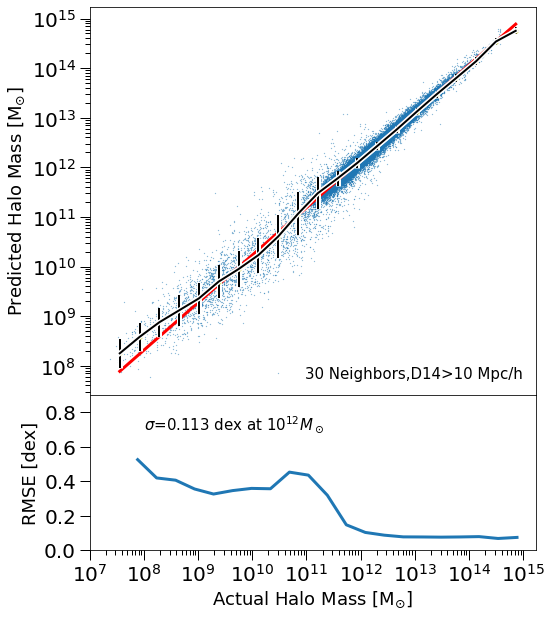}
        \caption{This figure shows a neural network trained only on halos that are more than $10$ Mpc/$h$ away from the nearest $10^{14}\Msun$ halo (similar to the Milky Way and M31, which are $\sim 11$ Mpc/$h$ away from the Virgo Cluster; \citealt{Mei07}). The uncertainties are very modestly lower (0.113 dex instead of 0.118} dex) at a halo mass of $10^{12}\Msun$.
        \label{fig:D14>10}
        \end{minipage}
\end{figure*}

To confirm our interpretation that neighbors of low-mass satellites are not providing any information about host halo masses, we trained a network just with three parameters ($D_{14}$, $D_\mathrm{larger}$ and $v_\mathrm{max,sat}$), and found similar errors for low-mass halos as compared to the network provided with full information about satellites (Figure~\ref{fig:3feat}). At the same time, the errors from this network (> 0.27 dex) imply that, for halos with $M_\mathrm{vir}>10^{11.7}\Msun$, adding orbital information for neighboring halos reduces the variance in predicted masses by $>90\%$. Hence, although $v_\mathrm{max,sat}$ is helpful to establish a broad prior on host halo mass, most of the information leading to the final predicted mass for MW-mass and larger halos is coming from neighboring halos' orbits.

Since we have shown that nearby massive halos impact neighboring halos' orbital distributions, we also consider a network trained on isolated halos (Fig.\ \ref{fig:D14>10}). Since the Milky Way and M31 are $\sim 11$ Mpc/$h$ from the Virgo Cluster \citep[e.g.,][]{Mei07}, we trained a separate network using only halos with $D_{14}>10$ Mpc/$h$. This network performed only marginally better (0.113 dex vs.0.118 dex errors for $10^{12}\Msun$ halos) than the network with no selection on $D_{14}$, suggesting that the network with no selection is nonetheless able to compensate well for the presence of a larger nearby halo.

\section{Discussion and Conclusions}

\label{s:discussion}

We find that applying a neural network with information from neighboring halo orbits can place tight constraints on the masses of Milky Way-like halos, with typical errors less than 0.12 dex. In our analysis, using information from 30 neighboring galaxies yields more accurate predictions of central halo masses compared to using only 10 neighboring galaxies, for which the uncertainties rise to $\sim $ 0.17 dex. This finding is consistent with the result reported by \citet{Patel18}, in that incorporating specific angular momenta as input variables allows for tight constraints in predicting central halo masses. 

Our approach offers several advantages over previous methods, addressing certain limitations and paving the way for future advancements. First, we have shown that it is not necessary to assume dynamical equilibrium or to assume satellite status to achieve tight constraints on halo masses, at least for halos with enough nearby satellites. Secondly, past simulation-based methods, such as those employed in \citet{Patel18} and others' previous works, may have slightly underestimated errors due to correlations between satellite orbits, regardless of whether the measurement errors are included or not. In our case, we find that going from 10 satellites to 30 satellites gives a factor of $\sqrt{2}$ improvement in uncertainties, whereas Poisson statistics would suggest a factor of $\sqrt{3}$. Part of the barrier in achieving lower (Poisson-limited) uncertainties could be due to correlations between satellite orbits, such as satellites arriving along the same filament. However, part of the barrier could also be limitations in characterizing the environment. For example, we showed that nearby high-mass halos cause contamination in satellite orbits, but other aspects of the environment could correlate with satellite orbits in as yet unexplored/unknown ways.

This study did not investigate the impact of observational errors beyond the fiducial observational error model in Appendix \ref{a:errors}, in part because we wished to understand the maximal amount of information present in satellite orbits. For a study that is applicable to the Milky Way and/or M31 systems, one would need to account for observational errors that correlate with heliocentric distance and other factors. This is the next planned step in our paper series, which will involve training a neural network on simulations with realistic observational errors and then using the resulting network to measure the masses of the Milky Way and Andromeda. Furthermore, our current work, similar to many previous studies, did not extensively test the method on hydrodynamical simulations. We recognize the importance of investigating the effectiveness of our approach on non-dark matter-only simulations, and we also plan to perform such tests. In particular, we plan to cross-validate the method by training on one hydrodynamical simulation and testing on another hydrodynamical simulation with a different physics implementation.

Beyond halo mass, we also plan to train new neural networks to estimate additional parameters such as the halo's spin axis, concentration, and assembly history. This would provide important context to our understanding of our own halo, including orbit modeling for satellites, as present halo models tend to assume a static mass and concentration history for the Milky Way. 

\section*{Acknowledgments}

EH and PB were funded through a Fellowship from the Packard Foundation, Grant \#2019-69646. EP acknowledges financial support provided by a grant for \textit{HST} archival program AR-16628 
through the Space Telescope Science Institute (STScI). EP also acknowledges financial support provided by NASA through the Hubble Fellowship grant \# HST-HF2-51540.001-A awarded by STScI. STScI is operated by the Association of Universities for Research in Astronomy, Incorporated, under NASA contract NAS5-26555.

 This research is based upon High Performance Computing (HPC) resources supported by the University of Arizona TRIF, UITS, and Research, Innovation, and Impact (RII) and maintained by the UArizona Research Technologies department. The University of Arizona sits on the original homelands of Indigenous Peoples (including the Tohono O’odham and the Pascua Yaqui) who have stewarded the Land since time immemorial.

EH thanks the LSSTC Data Science Fellowship Program, which is funded by LSSTC, NSF Cybertraining Grant \#1829740, the Brinson Foundation, and the Moore Foundation; her participation in the program has benefited this work.

The VSMDPL simulation was performed by Gustavo Yepes on the SuperMUC supercomputer at LRZ (Leibniz-Rechenzentrum) using time granted by PRACE, project number 012060963 (PI Stefan Gottloeber).

\bibliographystyle{mnras}
\bibliography{mnras_template} 
\appendix
\section{Fiducial Observational Uncertainties}

\label{a:errors}
The addition of observational uncertainties depends greatly on the system to which it is applied. For example, the relationship between observational errors and distance would be very different for satellites of the Milky Way compared to satellites of M31. However, to provide an order-of-magnitude estimate of how observational errors would affect our results, we incorporated approximate Milky Way-like errors (based on uncertainties in \citealt{2018AAS...23240203P,2021A&A...649A...7G}) into our simulations and re-trained neural networks on the perturbed data. Specifically, we added the following uncertainties to the parameters considered in our neural net:

\begin{itemize}
    \item $20\%$ relative error to the angular momentum of satellites,
    \item $10\%$ relative error to the  distance between satellites and their central halos,
    \item 30 km s$^{-1}$ to  satellite velocities,
    \item $5\%$ relative error to the distance to the nearest larger halo,
    \item $5\%$ relative error to the distance to the nearest $10^{14} \Msun$ halo, and
    \item $10\%$ error to the $V_\mathrm{max}$ of the most massive satellite.
\end{itemize}

Our expected performance is very similar when these approximate observational errors are included (it increased from 0.118 dex to 0.135 dex at $10^{12}\Msun$), which could be due to multiple possibilities.  One is that the expected observational errors are small relative to the intrinsic scatter in satellite properties across different halos, and another is that correlations between satellite orbits are partially mitigating the effect of scatter (i.e., in that the same information is present in multiple satellites, and so is more robust to the presence of noise).  The fact that the relative errors increased more for the 10-neighbor network ($\sim 20\%$) compared to the 30-neighbor network ($\sim 14\%$) suggests that this latter effect is present to some extent.  However, the fact that the increase is relatively low for both networks suggests that the observational errors are small relative to halo-to-halo dispersion.  This result gives us confidence that we can achieve a performance comparable to our model's ideal performance when applied to real data.

Additionally, it is important to highlight that the properties selected for our analysis ($j,V,R$) show weak correlations with the position or sky location in relation to the Milky Way.  This is important because real observations of satellites do not have consistent completeness across the full sky.  We would expect that the distribution of the chosen properties would not depend on sky coverage, but a full test of this would require a careful combination of many different surveys' completeness maps for the Milky Way, beyond the scope of this paper.
\begin{figure*}
  \begin{minipage}{0.5\textwidth}
    \centering
    \includegraphics[width=\linewidth]{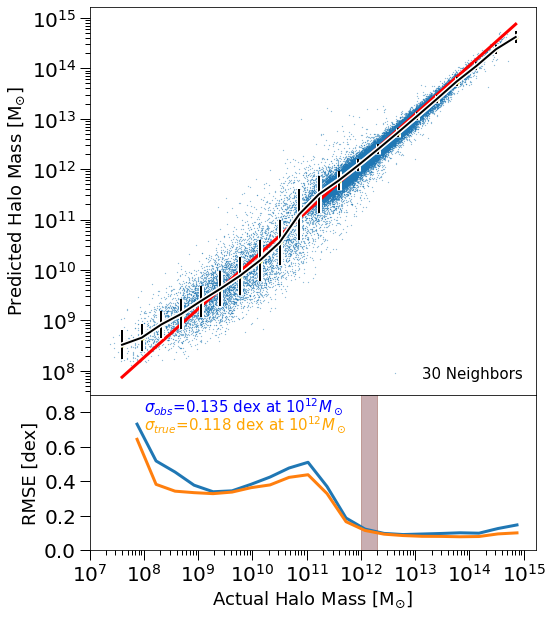}
  \end{minipage}%
  \begin{minipage}{0.5\textwidth}
    \centering
    \includegraphics[width=\linewidth]{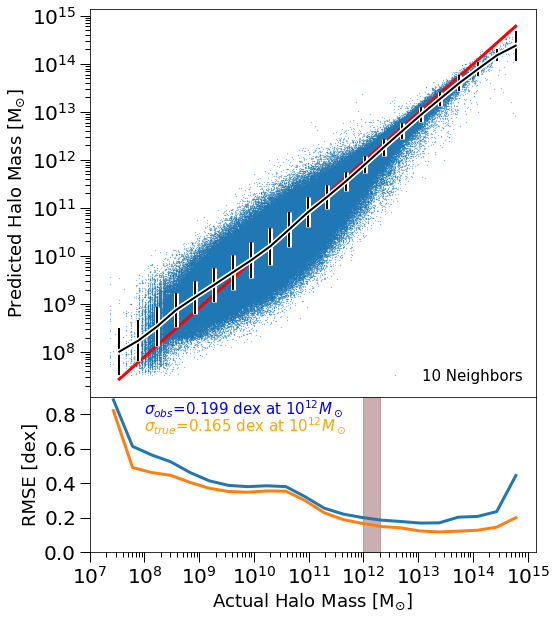}
  \end{minipage}
  \caption{Analogous to Figure~\ref{fig:overall}, these figures show the predicted halo mass versus actual halo mass when adding reasonable observational errors to our model. The input features fed into the networks correspond to observables, primarily focusing on the specific angular momenta and other orbital properties of neighboring galaxies. In the comparison between the left and right figures, the left figure displays outcomes from halos with a minimum of 30 neighboring galaxies, exhibiting reduced errors compared to the right figure, which considers halos with a minimum of 10 neighboring galaxies. The bottom panels illustrate the root mean square error (RMSE) versus the actual halo mass. The blue curve represents the addition of fiducial observational errors, while the simulation data without any observational errors is depicted by the orange curve. Notably, the introduction of fiducial observational errors does not substantially alter the results. For networks utilizing 10 neighbors, typical errors amount to approximately 0.199 dex for Milky Way-mass halos, whereas for networks utilizing 30 neighbors, these errors are about 0.135 dex. Error bars depict the standard deviations of predicted halo masses relative to actual halo masses, while the black line indicates the medians of predicted halo masses within bins of actual halo masses. The gray bar at $10^{12}\Msun$ corresponds to the approximate mass of the Milky Way. Finally, the red line serves as a reference, indicating where the predicted mass would align with the actual mass.}
  \label{fig:error}
\end{figure*}








\end{document}